# Optical Control of Topological Polariton Phase in a Perovskite Lattice

Rui Su[1,4]*, Sanjib Ghosh[1,4], Timothy C.H. Liew[1,2]* & Qihua Xiong[1,2,3]*

[1]Division of Physics and Applied Physics, School of Physical and Mathematical Sciences, Nanyang Technological University, 637371, Singapore.

[2]MajuLab, International Joint Research Unit UMI 3654, CNRS, Université Côte d'Azur, Sorbonne Université, National University of Singapore, Nanyang Technological University, Singapore.

[3]State Key Laboratory of Low-Dimensional Quantum Physics and Department of Physics, Tsinghua University, Beijing, China

[4]These authors contributed equally to this work.

*Corresponding author. Email:Qihua_xiong@tsinghua.edu.cn (Q.X.); Timothyliew@ntu.edu.sg (T.L.) ; surui@ntu.edu.sg (R.S.)

**Strong light-matter interaction enriches topological photonics by dressing light with matter, which provides the possibility to realize tuneable topological devices with immunity to defects. Topological exciton polaritons, half-light half-matter quasiparticles with giant optical nonlinearity represent a unique platform for active topological photonics with phase tunability. Previous demonstrations of exciton polariton topological insulators still demand cryogenic temperatures and their topological properties are usually fixed without phase tunability. Here, we experimentally demonstrate a room-temperature exciton polariton topological insulator with active phase tunability in a perovskite zigzag lattice. Polarization serves as a degree of freedom to control the reversible transition between distinct topological phases, thanks to the polarization-dependent anisotropy in halide perovskite microcavities. The topologically nontrivial polariton states localized in the edges persist in the presence of a natural defect, showing strong immunity to disorder. We further demonstrate that exciton polaritons can condense into the topological edge states under optical pumping. These results provide an ideal platform for realizing tuneable topological polaritonic devices with room-temperature operation, which can find important applications in optical control, modulation and switching**.

Topological insulators represent a new class of materials that are insulating in the bulk, but show the hallmark property of topological conducting edge states, which are endowed with topological robustness against disorder and structural imperfections, as a consequence of the nontrivial topology of the bulk[1]. Early demonstrations of topological insulators date back to the discovery of quantum Hall effect in condensed matter in 1980s[2] and the concept of topology was later advanced to various fields including microwaves[3,4], photonics[5,6], cold atoms[7], acoustics[8], and also mechanics[9]. Particularly, initially proposed by Haldane and Raghu[3], the extension of topology into photonics has revolutionized the design principle of novel optical devices, which allow robust edge transport with strongly suppressed backscattering loss[4,10,11], robust lasing[12-14] and harmonic generation[15], as well as robust propagation of single photons in the quantum regime[16,17]. Despite the remarkable success of introducing topology into photonics for robustness, the properties of the topological photonic structures are usually fixed once fabricated. To gain more freedom of control for a broader range of applications, tuneable topological insulators are highly demanded[18-21], such that the topological character could be controlled on-demand to be on or off. In condensed matter, several strategies have been experimentally utilized to tune the topological phase via spin orbit coupling[22] and electrical field[23], which lay the foundation for realizing topological transistors. Similar manipulations have also been experimentally demonstrated with microwaves[18], mechanics[24] and electric circuits[25,26]. However, such on-demand control remains experimentally nontrivial in the optical frequency, particularly the visible spectral range, which demands subwavelength architectures[27].

Recently, microcavity exciton polariton system emerges as a unique platform for topological photonics[11,12,28-34], representing a linking bridge between photonics and condensed matter. Exciton polaritons are half-light, half matter quasiparticles, resulted from the strong coupling between cavity photons and excitons. Compared with pure photonic systems, the light-matter hybridization provides them strong nonlinearity and enhanced response to external stimuli, which provides the possibility to control the topological phases. Topological exciton polaritons have been experimentally demonstrated in GaAs microcavities[11,12] and the more recent monolayer WS$_2$[33], but demanded cryogenic temperatures and lacked phase tuneability. With the recent demonstration of polariton condensation in lattices[35,36], the halide perovskite system[37,38] emerges as a promising candidate for room temperature active topological insulators. Exciton polariton topological insulators with active phase tunability were theoretically proposed before[30], while they have not been experimentally realized yet.

Here, we demonstrate a exciton polariton topological insulator with active phase tunability in one-dimensional perovskite lattice at room temperature, which is based on the Su–Schrieffer–Heeger (SSH) model by coupling the *s*-orbital type polariton modes with a zigzag chain of nanopillars. By employing the anisotropy and strong photonic spin-orbit coupling from the halide perovskite microcavity[35], we demonstrate the emergence of topological polariton edge states locating inside a large topological gap of ~10 meV, suggesting a stronger immunity against defects than previous observations[11,12]. A large topological gap is also essential for nonlinear devices that seek to use energy shifts to realize switches and information processing while remaining topological[32]. In the meantime, the polariton zigzag lattice can be tuned on-demand between the topologically nontrivial phase and the trivial phase by means of polarization control. Furthermore, we show the topological robustness of the polariton edge states in the presence of a natural defect and exciton polaritons can condense into the topological edge states under optical pumping at room temperature.

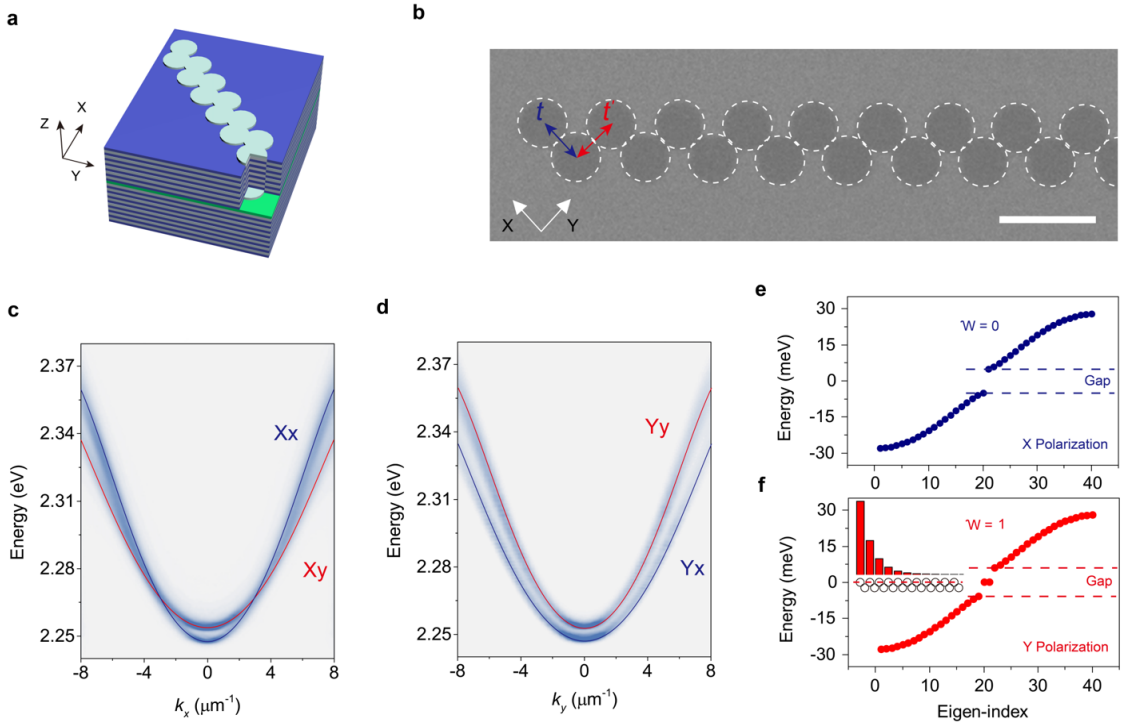

**Fig. 1**. **Schematic and mechanism of polarization-dependent topological phase in a perovskite zigzag lattice. a,** Schematic diagram of the perovskite zigzag lattice, where the lattice is created by patterning the PMMA spacer layer and aligned 45 degree along the Y axis of perovskite layer. **b,** Scanning electron microscopy image of the zigzag chain on perovskite layer before the deposition of top DBR. The white dashed circles are added for visibility. *t* and *t'* represent the intracell and intercell hopping strengths. Scale bar, 1 μm. **c**, **d**, Polariton dispersions collected along $k_x$ at $k_y = 0\ \mu m^{-1}$ and along $k_y$ at $k_x = 0\ \mu m^{-1}$ from a planar perovskite microcavity under X and Y polarizations at room temperature, respectively. The solid curves represent the fittings from a coupled oscillator model for X (Blue) and Y polarization (Red). **e**, **f**, Calculated band structures in X and Y polarizations, respectively. Y polarization represent a topological nontrivial phase ($W=1$) with topological polariton edge states inside the topological gap while X polarization represent a trivial phase ($W=0$). Inset to **f**, The wavefunction distribution of the polariton edge state along the pillars highlight with red dashed line, showing an exponential decay of $(t_Y/t_Y')^n$.

In our experiment, we fabricate an active SSH zigzag chain consisting of 40 identical coupled nanopillars (20 unit cells) with a diameter of 0.5 μm and center-to-center distance of 0.475 μm, which is schematically shown in Figure 1a. The zigzag chain is created by patterning the spacer layer of poly(methyl methacrylate) (PMMA) inside the cesium lead bromide perovskite planar microcavity (Figure 1b) with a cavity detuning of ~140 meV. Unlike previous demonstrations of topology by employing *p*-orbital type polariton modes with a limited topological gap and no phase tunability at 4 K[12], here we create the topology with *s*-orbital type polariton modes with a substantially large topological gap and active phase tunability at room temperature, as a consequence of the intrinsic anisotropy and strong photonic spin-orbit coupling of perovskite system. As shown in Figure 1a and Supplementary Part 1, we intentionally align the zigzag chain at an angle of 45 degree with respect to one of the perovskite crystalline axes. In this case, the polariton hopping directions inside the lattice are exactly along the major crystalline axes. In the SSH model, the unit cell is composed of two sites and the Hamiltonian can be written in the tight binding limit as

$$\hat{H} = \sum_n (t\hat{a}_n\hat{b}_n^\dagger + t'\hat{a}_n^\dagger\hat{b}_n + h.c.),$$

where $\hat{a}_n(\hat{a}_n^\dagger)$ and $\hat{b}_n(\hat{b}_n^\dagger)$ are the polariton creation (annihilation) operators in the sublattices *a* and *b* in the *n*th unit cell, respectively; *t* and *t'* represent the intracell and intercell hopping strengths, respectively. The Hamiltonian reveals two topologically distinct phases for the cases of *t*>*t'* (trivial) and *t*<*t'* (topologically nontrivial), and the difference can be directly distinguished by the topological invariant, *i.e.*, the winding number $W$ of phase $\phi(k)$ across the Brillouin zone,

$$W = \frac{1}{2\pi} \int_{BZ} \frac{\partial \phi(k)}{\partial k} dk$$

$W=1$ for the topologically nontrivial phase (*t*<*t'*) and $W=0$ for the trivial phase (*t*>*t'*). In our perovskite system, due to the intrinsic anisotropy and strong photonic spin-orbit coupling[35], the polariton effective mass is strongly polarization-dependent and direction-dependent. As shown in Figure 1c and 1d, the dispersions from a perovskite planar microcavity exhibit distinct curvatures under different polarizations and directions (along X and Y), suggesting different effective masses that link to the polariton hopping strengths (Methods). From the fitting of the dispersions, we extract distinct effective masses of $m_{xx} = 1.5188 \times 10^{-5} m_e$ (X direction in X polarization, blue line in Figure 1c), $m_{xy} = 2.2335 \times 10^{-5} m_e$ (X direction in Y polarization, red line in Figure 1c), $m_{yy} = 1.5188 \times 10^{-5} m_e$ (Y direction in Y polarization, red line in Figure 1d), and $m_{yx} = 2.1094 \times 10^{-5} m_e$ (Y direction in X polarization, blue line in Figure 1d), respectively. In this scenario, the hopping strength ratios between intracell and intercell appear to be different with X and Y polarizations, where $t_Y/t_Y' = 0.68$ for Y polarization and $t_X/t_X' = 1.39$ for X polarization. This suggests that the system exhibits a topologically nontrivial phase with Y polarization ($W = 1$) while shows the trivial phase with X polarization ($W = 0$). Figure 1e and Figure 1f display the calculated band structures for a zigzag chain of 40 sites with X and Y polarizations. Topological gaps open in both X and Y polarizations, while the most notable difference is that topological states exist in the latter case inside the topological gap and spatially localize at the two end sites of the zigzag chain. The localization nature is strongly determined by the hopping strength ratio, where the calculated wavefunction distribution of the topological state exhibits an exponential decay behavior of $(t_Y/t_Y')^n$ along the pillars as highlighted in the inset of Figure 1f.

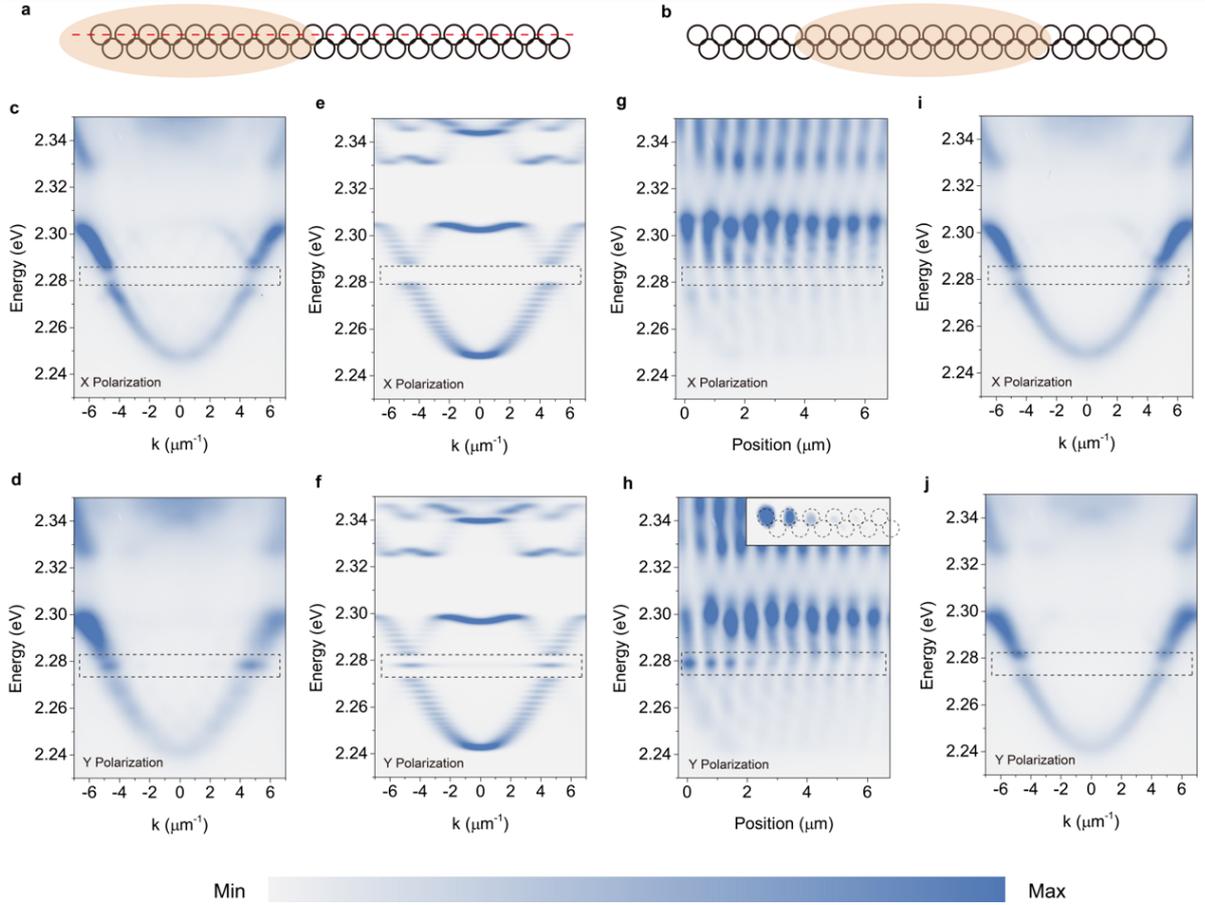

**Fig. 2**. **Experimental demonstrations of polarization-dependent phases in a perovskite zigzag lattice. a**, **b,** Schematic diagrams of excitation and emission collection from the edge and the body center, covering 10 unit cells, respectively. The red dashed line represents the emission collection area for the energy-resolved spatial images in **g** and **h**. **c**, **d**, Energy-resolved momentum-space polariton dispersions of the perovskite zigzag lattice collected from the edge in **a** for X and Y polarizations, respectively. The grey dashed lines highlight the topological gap opening inside the *s* band. Topological polariton edge states only exist inside the topological gap in Y polarization. **e, f,** Theoretically calculated momentum-space polariton dispersions collected from the edge for X and Y polarizations, respectively. **g, h,** Energy-resolved spatial images collected along the red dashed line in **a** for X and Y polarizations, respectively. It shows localized emission at the end site with some decay to nearby pillars in Y polarization. Inset to **h**, a typical real space image at the edge energy, showing localized emission at the end site. **i, j,** Energy-resolved momentum-space polariton dispersions of the perovskite zigzag lattice collected from the body center in **b** for X and Y polarizations, respectively.

In order to experimentally demonstrate the emergence of topological polariton edge states and active control of distinct phases, we directly probe the band structure of the zigzag chain by mapping the energy-resolved momentum-space photoluminescence spectra. The zigzag chain is non-resonantly excited with a continuous-wave laser of 457 nm (2.713 eV) at room temperature. Figure 2a and 2b display the emission collection schemes from the edge and body center of the lattice, which cover around 10 unit cells, respectively. Figure 2c shows the experimental photoluminescence mapping in X polarization collected from the edge. Due to the subwavelength size of the nanopillars, the coupling between *s*-mode polaritons gives rise to the *s* band ranging from 2.24 eV to 2.30 eV, representing the major band in the dispersions. At higher energies above 2.32 eV, the *p* band arising from *p*-mode polaritons can also be seen. Within the *s* band, the anisotropic hopping of polaritons results in a large topological gap opening of ~8 meV as highlighted with grey dashed line, which is in good agreement with the theoretically calculated dispersion shown in Figure 2e. The spatial image collected along the red dashed line (Fig. 2a) further elucidates the topological gap opening nature, where no emission is observed inside the topological gap but the emission from the pillars can be observed outside the topological gap within *s* band. As a striking comparison, the emission collected from the edge in Y polarization (Fig. 2d) exhibits distinct features, where a similar topological gap opening of ~10 meV exists inside the *s* band but with the additional emergence of discrete states inside the topological gap at $k = 4.5 \ \mu m^{-1}$ (highlighted with grey dashed line). Such discrete states correspond to the emergence of topological polariton edge states in Y polarization, which is in good agreement with theory (Fig. 1f and Fig. 2f). The large topological gap opening of ~10 meV is at least 5 times larger than previous demonstrations with polaritons[11,12], which ensures a higher immunity to external perturbations. The nontrivial nature can be further experimentally clarified by the real space image at the edge state energy and energy-resolved spatial image (Fig. 2h), which exhibit strongly localized emission at the end site. Due to the limited difference between intracell and intercell hopping strength ($t_Y / t_Y' = 0.68$), the edge state wavefunction exhibits some decay to nearby pillars along the red dashed line (Fig. 2a). To further confirm the emergence of topological states localized at the edge, as shown in Figure 2i and Figure 2j, the energy-resolved emission mappings collected from the body center (Fig. 2b) exhibit similar topological gaps inside the *s* band but without any discrete states inside the topological gap, in both X and Y polarizations. These results together unambiguously suggest that the perovskite zigzag lattice works as a trivial insulator in X polarization but a topologically nontrivial insulator in Y polarization, which can be actively tuned on-demand by polarization control. Such realization of active tunability with a substantial topological gap provides an ideal platform to explore active and tuneable topological devices with strong immunity at room temperature.

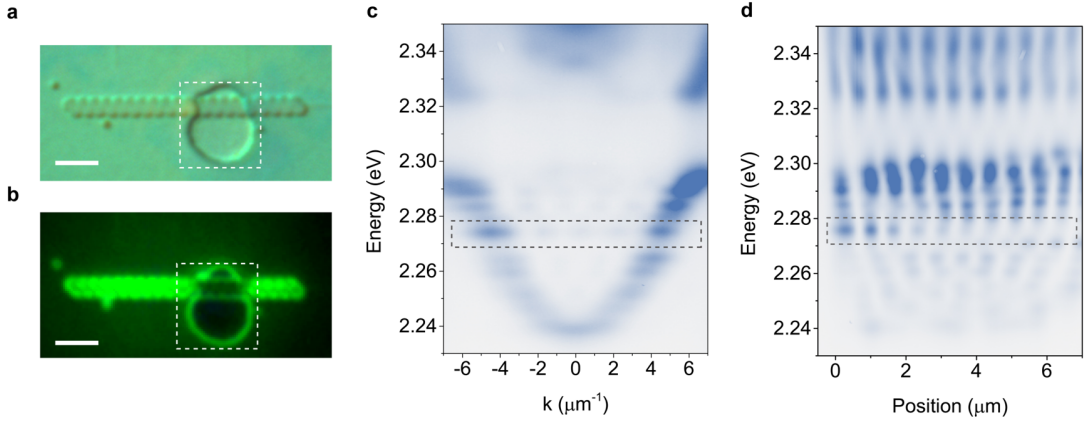

**Fig. 3**. **Topological edge states in the presence of a natural defect. a**, **b**, Optical microscopy and fluorescence images of the zigzag lattice aligned in a hole of perovskite layer. The white dashed lines highlight the hole of the perovskite layer. Scale bar, 2 μm. **c**, **d**, Energy-resolved momentum-space polariton dispersions and spatial images of the perovskite zigzag lattice in Y polarization in the presence of structure imperfections. The grey dashed lines highlight the topological gap opening with topological edge states inside the *s* band. The topological polariton edge states persist in the presence of a natural defect, showing the robustness.

One of the most crucial features of topologically nontrivial states is their robustness against fabrication imperfections and deformations, which provide the promising future in real world applications. Here we demonstrate such robustness with a natural defect. Specifically, as shown in Figure 3a, a hole on the perovskite active medium is naturally induced during the perovskite transfer process and we align part of the zigzag chain inside the hole (highlighted with white dashed line). The fluorescence image of the zigzag chain further elucidates the missing part of perovskite in the hole, as shown in Figure 3b, where we observe no emission from the hole. In this scenario, the lattice potential inside the hole changes compared with that of outside the hole, due to the lack of perovskite active medium, which acts as the natural defect of the lattice. As shown in Figure 3b, topological polariton edge states remain inside the energy gap in Y polarization with localized emission from the last pillar at the left end (Fig. 3d). While in X polarization, only topological gap can be observed with no discrete states inside from both energy-resolved momentum-space and spatial photoluminescence mappings (Supplementary Part 2). This evidence shows that polarization-controlled phases between topologically nontrivial and trivial persist in the presence of fabrication imperfections and deformations.

An interesting feature for exciton polaritons in the strong coupling regime is that they are able to show non-equilibrium condensation at high temperatures, accompanied by laser-like emission. Thanks to the driven-dissipative nature of polaritons, in which the steady state is fixed by the interplay between gain and loss, polaritons can condense to non-ground states[11,12,38], which has been shown in perovskite lattices recently[35]. In order to achieve polariton condensation in the topological edge states, we select a sample with exciton photon detuning of ~ -140 meV, in which polaritons favour relaxation into the edge state mode. The system is non-resonantly pumped with a pulsed laser to reach the condensation regime. As shown in Figure 4a, under low excitation power of 0.5 $P_{th}$, where $P_{th}$ is the threshold pump fluence, the perovskite lattice exhibits a typical dispersion in Y polarization with edge states in accordance with Fig. 2g, while under strong excitation power of 2.0 $P_{th}$, polaritons tends to condense into selected states with maximum gain. As shown in Fig. 4b, the emission is mainly from the edge states at $k = 4.5\ \mu m^{-1}$ with much more intense intensity, suggesting the emergence of condensation into the polariton edge states. Polariton edge state condensation could be further revealed by the real space image (Fig. 4d), where it shows localized emission at the end site with some decay to nearby sites. It is worth to note that we also observe much weaker emission from a higher state, corresponding to the state at the top of *s* band at $k = 6.2\ \mu m^{-1}$, which could be confirmed from the wavevector and also the real space image (Fig. 4c). The simultaneous condensation into this state could be due to the overlapping of the large pump spot with the bulk mode, which provides the gain to condense. To characterize the transition quantitatively, we demonstrate the evolution of emission intensity, linewidth and peak energy at the edge state as a function of pump fluence. As shown in Figure 4e, with the increase of pump fluence, the emission from the edge state exhibits a clear nonlinear increase by three orders with a threshold of $P_{th} = 10\ \mu J/cm^2$. Correlatively, the linewidth of the polariton edge state displays a clear drop by around three times from 7.2 meV to 2.6 meV across the threshold. In the meantime, the energy of the polariton edge state emission exhibits a clear continuous blueshift trend with increasing the pump fluence (Fig. 4f), as a consequence of repulsive interaction of polaritons and reservoir. All the above evidence together shows the emergence of polariton condensation into the edge states.

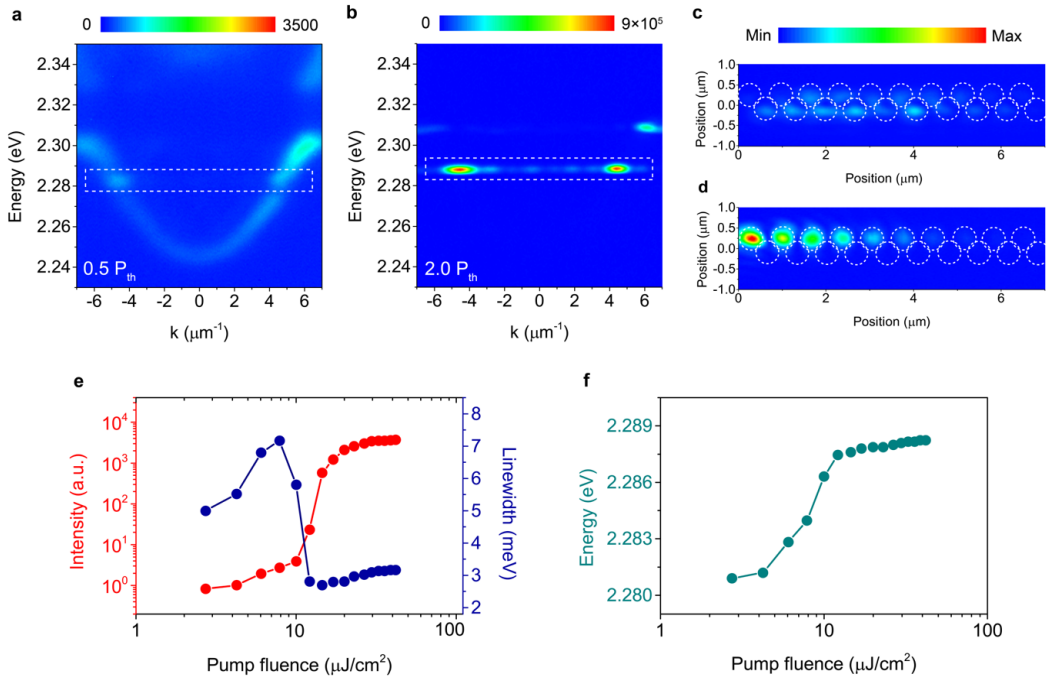

**Fig. 4**. **Polariton Condensation into the topological edge state at room temperature. a,** Energy-resolved momentum-space polariton dispersion at 0.5 $P_{th}$ in Y polarization, showing a typical dispersion with discrete states at $k = 4.5\ \mu m^{-1}$ inside the topological gap. The white dashed lines highlight the topological gap with topological polariton edge states inside the *s* band. **b,** Energy-resolved momentum-space polariton dispersion at 2.0 $P_{th}$ in Y polarization,

showing dominant emission from the edge states at $k = 4.5\ \mu m^{-1}$. Simultaneous condensation into a higher energy state with much weak emission is also observed at $k = 6.2\ \mu m^{-1}$, suggesting the driven-dissipative nature of polaritons. **c, d**, Real space images of higher energy state and the topological edge states at 2.0 $P_{th}$. The real space image of the state at $k = 4.5\ \mu m^{-1}$ exhibit strong localized emission at the end site with some decay to nearby sites, suggesting the edge state nature. **e**, Evolution of integrated output intensity and linewidth of the edge states as a function of the pump fluence, which exhibit a clear intensity increase by three orders and linewidth drop by around three times across the threshold of $P_{th} = 10\ \mu J/cm^2$. **f**. Emission energy evolution of the edge states as a function of the pump fluence, showing a continuous blueshift trend as a consequence of repulsive interactions.

In conclusion, we have unambiguously demonstrated a one-dimensional exciton polariton topological insulator with active phase tuneability at room temperature. By employing the anisotropy and strong photonic spin-orbit coupling of halide perovskite microcavities, we demonstrated polarization-dependent topological phases in a zigzag chain, which could be tuned on-demand. Such tuneability persists in the presence of natural structural imperfections, showing the robustness associated to topology. In addition, we also demonstrate that exciton polaritons could condense to this particular topological edge state with optical pumping at room temperature. Our work reveals a new avenue to tune the topological phases by photonic spin-orbit coupling and we anticipate that more tunability degrees of freedom can be achieved with synthetic photonic spin-orbit coupling[39]. Our results represent an important step not only towards explore fundamental topological polaritonics, but also towards developing tuneable optical devices with robustness.

# Methods

**Perovskite Lattice Fabrication.** 20.5 pairs of titanium oxide and silicon dioxide was deposited using an electron beam evaporator as the bottom DBR. The 80 nm-thick cesium lead bromide perovskite was grown with a vapor phase deposition on a mica substrate and transferred onto the bottom DBR by a dry-transfer process with scotch tape. A thin layer of PMMA spacer with thickness of 60 nm was spincoated onto the perovskite layer and patterned into zigzag chain with a standard E beam lithography process. Another 10.5 pairs of silicon dioxide and tantalum pentoxide was deposited onto the structure by the e-beam evaporator, acting as the top DBR to complete the fabrication process.

**Optical spectroscopy characterizations.** The energy-resolved momentum-space and spatial photoluminescence mappings are conducted using a home-built angle resolved photoluminescence setup. A 50 × objective (NA= 0.75, Mitutoyo) was used to collect the emission from the perovskite lattice which was sent to a 550-mm focal length spectrometer (HORIBA iHR550) with a grating of 600 lines/mm and a liquid nitrogen–cooled charge coupled device of 256×1024 pixels. In the linear region, a continuous-wave laser (457 nm) with a pump spot of ~10 μm was used to pump the perovskite lattice. In the real space plane, a spatial filter was used to select the emission area of the lattice. In the nonlinear regime, the perovskite lattice was pumped non-resonantly with a pulsed laser (400 nm, pulse duration of 100 fs, repetition rate of 1 kHz) with a pump spot of ~15 μm.

**Theoretical Calculations.** We consider an anisotropic model for photons to describe the polaritons in perovskite. The anisotropy is considered in the form of different effective masses along two perpendicular directions to the crystal axis.

$$i\hbar \frac{\partial \varphi}{\partial t}(\vec{r},t) = \left[ -\frac{\hbar^2}{2m_y}\frac{\partial^2}{\partial y^2} - \frac{\hbar^2}{2m_x}\frac{\partial^2}{\partial x^2} + V(\vec{r}) \right] \varphi(\vec{r},t) + \frac{g_0}{2}\chi(\vec{r},t) \quad (1)$$

Photons couple with excitons to form polaritons. We describe the excitons with the wave function $\chi(\vec{r})$ which follows the equation:

$$i\hbar \frac{\partial \chi}{\partial t}(\vec{r},t) = E_{ex}\chi(\vec{r},t) + \frac{g_0}{2}\varphi(\vec{r},t) \quad (2)$$

The potential $V(\vec{r})$ is introduced for the photons to form a zigzag lattice potential (Supplementary Part 3). It is experimentally shown that the masses $m_x$ and $m_y$ are polarization dependent. Let us rewrite the coupled dynamical equation for photons and excitons as

$$i\hbar \frac{\partial}{\partial t}\begin{pmatrix} \varphi(\vec{r}) \\ \chi(\vec{r}) \end{pmatrix} = \begin{pmatrix} \hat{H}p & g_0/2 \\ g_0/2 & E_{ex} \end{pmatrix}\begin{pmatrix} \varphi(\vec{r}) \\ \chi(\vec{r}) \end{pmatrix} \quad (3)$$

Where the photon Hamiltonian is given by $\hat{H}p = -\hbar^2/(2m_y)\partial_y^2 - \hbar^2/(2m_x)\partial_x^2 + V(\vec{r})$. The potential $V(\vec{r})$ defines the lattice as the zigzag chain (Supplementary Part 3). We diagonalize the Hamiltonian to obtain the modes $\varphi_n$ and the energy eigenvalues $\varepsilon_n$. We calculate the dispersion from the formula,

$$I(E,\vec{k}) = \frac{1}{\pi\Delta_k\Delta_E}\sum_{n,\vec{q}}|\tilde{\varphi}_n(\vec{q})|^2 \exp\left[-\frac{(E-\varepsilon_n)^2}{\Delta_E^2} - \frac{|\vec{k}-\vec{q})^2|}{\Delta_k^2}\right] \qquad (4)$$

where $\Delta_E$ and $\Delta_k$ are the energy and momentum resolutions in the dispersion. $\tilde{\varphi}_n(\vec{q})$ is the Fourier transform of the eigenfunction $\varphi_n(r)$. Experimentally, we find that for one polarisation $m_x/m_y>1$ while for the other one it is opposite $m_x/m_y<1$. For our numerical simulation we considered $E_{ex}$ = 2407.7 meV and $g_0$ = 120 meV. For Topological phase in Y polarization: $m_y/m_x = 0.68$, $m_x = 2.2335\times10^{-5} m_e$, $m_y = 1.5188\times10^{-5} m_e$. For trivial phase in X polarization: $m_y/m_x=1.3889$, $m_x = 1.5188\times10^{-5} m_e$, $m_y = 2.1094\times10^{-5} m_e$ ($m_e$ is the electron mass).

**References:**

1. Hasan, M. Z. & Kane, C. L. Colloquium: Topological insulators. *Rev. Mod. Phys.* **82**, 3045-3067, (2010).
2. Klitzing, K. v., Dorda, G. & Pepper, M. New Method for High-Accuracy Determination of the Fine-Structure Constant Based on Quantized Hall Resistance. *Phys. Rev. Lett.* **45**, 494-497, (1980).
3. Haldane, F. D. M. & Raghu, S. Possible Realization of Directional Optical Waveguides in Photonic Crystals with Broken Time-Reversal Symmetry. *Phys. Rev. Lett.* **100**, 013904, (2008).
4. Wang, Z., Chong, Y., Joannopoulos, J. D. & Soljačić, M. Observation of unidirectional backscattering-immune topological electromagnetic states. *Nature* **461**, 772-775, (2009).
5. Rechtsman, M. C. *et al.* Photonic Floquet topological insulators. *Nature* **496**, 196-200, (2013).
6. Hafezi, M., Mittal, S., Fan, J., Migdall, A. & Taylor, J. M. Imaging topological edge states in silicon photonics. *Nat. Photon.* **7**, 1001-1005, (2013).
7. Jotzu, G. *et al.* Experimental realization of the topological Haldane model with ultracold fermions. *Nature* **515**, 237-240, (2014).
8. Yang, Z. *et al.* Topological Acoustics. *Phys. Rev. Lett.* **114**, 114301, (2015).
9. Süsstrunk, R. & Huber, S. D. Observation of phononic helical edge states in a mechanical topological insulator. *Science* **349**, 47-50, (2015).
10. Khanikaev, A. B. *et al.* Photonic topological insulators. *Nat. Mater.* **12**, 233-239, (2013).
11. Klembt, S. *et al.* Exciton-polariton topological insulator. *Nature* **562**, 552-556, (2018).
12. St-Jean, P. *et al.* Lasing in topological edge states of a one-dimensional lattice. *Nat. Photon.* **11**, 651-656, (2017).
13. Bahari, B. *et al.* Nonreciprocal lasing in topological cavities of arbitrary geometries. *Science* **358**, 636-640, (2017).
14. Bandres, M. A. *et al.* Topological insulator laser: Experiments. *Science* **359**, eaar4005, (2018).
15. Kruk, S. *et al.* Nonlinear light generation in topological nanostructures. *Nature Nanotechnology* **14**, 126-130, (2019).
16. Barik, S. *et al.* A topological quantum optics interface. *Science* **359**, 666-668, (2018).
17. Mittal, S., Goldschmidt, E. A. & Hafezi, M. A topological source of quantum light. *Nature* **561**, 502-506, (2018).
18. Cheng, X. *et al.* Robust reconfigurable electromagnetic pathways within a photonic topological insulator. *Nat. Mater.* **15**, 542-548, (2016).
19. Leykam, D., Mittal, S., Hafezi, M. & Chong, Y. D. Reconfigurable Topological Phases in Next-Nearest-Neighbor Coupled Resonator Lattices. *Phys. Rev. Lett.* **121**, 023901, (2018).
20. Jung, M., Fan, Z. & Shvets, G. Midinfrared Plasmonic Valleytronics in Metagate-Tuned Graphene. *Phys. Rev. Lett.* **121**, 086807, (2018).
21. Kudyshev, Z. A., Kildishev, A. V., Boltasseva, A. & Shalaev, V. M. Photonic topological phase transition on demand. *Nanophotonics* **8**, 1349, (2019).
22. Xu, S.-Y. *et al.* Topological Phase Transition and Texture Inversion in a Tunable Topological Insulator. *Science* **332**, 560-564, (2011).
23. Collins, J. L. *et al.* Electric-field-tuned topological phase transition in ultrathin Na3Bi. *Nature* **564**, 390-394, (2018).
24. Li, S., Zhao, D., Niu, H., Zhu, X. & Zang, J. Observation of elastic topological states in soft materials. *Nat. Commun.* **9**, 1370, (2018).
25. Hadad, Y., Soric, J. C., Khanikaev, A. B. & Alù, A. Self-induced topological protection in nonlinear circuit arrays. *Nature Electronics* **1**, 178-182, (2018).
26. Zangeneh-Nejad, F. & Fleury, R. Nonlinear Second-Order Topological Insulators. *Phys. Rev. Lett.* **123**, 053902, (2019).
27. Peng, S. *et al.* Probing the Band Structure of Topological Silicon Photonic Lattices in the Visible Spectrum. *Phys. Rev. Lett.* **122**, 117401, (2019).

28. Bardyn, C.-E., Karzig, T., Refael, G. & Liew, T. C. H. Topological polaritons and excitons in garden-variety systems. *Phys. Rev. B* **91**, 161413, (2015).
29. Karzig, T., Bardyn, C.-E., Lindner, N. H. & Refael, G. Topological Polaritons. *Phys. Rev. X* **5**, 031001, (2015).
30. Solnyshkov, D. D., Nalitov, A. V. & Malpuech, G. Kibble-Zurek Mechanism in Topologically Nontrivial Zigzag Chains of Polariton Micropillars. *Phys. Rev. Lett.* **116**, 046402, (2016).
31. Nalitov, A. V., Solnyshkov, D. D. & Malpuech, G. Polariton $\mathbb{Z}$ Topological Insulator. *Phys. Rev. Lett.* **114**, 116401, (2015).
32. Banerjee, R., Mandal, S. & Liew, T. C. H. Coupling between Exciton-Polariton Corner Modes through Edge States. *Phys. Rev. Lett.* **124**, 063901, (2020).
33. Liu, W. *et al.* Generation of helical topological exciton-polaritons. *Science*, eabc4975, (2020).
34. Pickup, L., Sigurdsson, H., Ruostekoski, J. & Lagoudakis, P. G. Synthetic band-structure engineering in polariton crystals with non-Hermitian topological phases. *Nat. Commun.* **11**, 4431, (2020).
35. Su, R. *et al.* Observation of exciton polariton condensation in a perovskite lattice at room temperature. *Nat. Phys.* **16**, 301-306, (2020).
36. Dusel, M. *et al.* Room temperature organic exciton–polariton condensate in a lattice. *Nat. Commun.* **11**, 2863, (2020).
37. Su, R. *et al.* Room-Temperature Polariton Lasing in All-Inorganic Perovskite Nanoplatelets. *Nano Lett.* **17**, 3982-3988, (2017).
38. Su, R. *et al.* Room temperature long-range coherent exciton polariton condensate flow in lead halide perovskites. *Sci. Adv.* **4**, eaau0244, (2018).
39. Rechcińska, K. *et al.* Engineering spin-orbit synthetic Hamiltonians in liquid-crystal optical cavities. *Science* **366**, 727-730, (2019).